\documentclass[twoside]{article}
\usepackage{epsfig.sty}
\newcommand{\beq}{\begin{eqnarray}}
\newcommand{\eeq}{\end{eqnarray}}
\begin{document}
\title{Z boson production via p-p and Pb-Pb collisions at $\sqrt{s_{pp}}$=5.02
  TeV}
\author{Leonard S. Kisslinger$^{1}$\\
Department of Physics, Carnegie Mellon University, Pittsburgh PA 15213 USA.\\
Debasish Das$^{2}$\\
Saha Institute of Nuclear Physics, HBNI 1/AF Bidhannagar Kolkata 700064 INDIA.\\
\hspace{1cm}(1) kissling@andrew.cmu.edu\hspace{1cm}(2) dev.deba@gmail.com }
\date{}
\maketitle

\begin{abstract}
   We estimate the production of $Z^a$ bosons via p-p collisions using 
previous work on $J/\Psi$, $\Psi(2S)$ production in p-p collisions, 
with the new aspect being the creation of $Z^a$ bosons via quark interactions,
with $a$ the component of a $Z$ vector boson. We then estimate the production
of $Z^a$ bosons via Pb-Pb collisions using modification factors from previous
publications.
\end{abstract}
\noindent
PACS Indices:12.38.Aw,13.60.Le,14.40.Lb,14.40Nd
\vspace{1mm}


\section{Introduction}

  This an extension of our recent work on heavy quark state production via
Xe-Xe collisions at $\sqrt{s_{pp}}$=5.44 TeV\cite{kd18} and heavy quark state 
production in Pb-Pb collisions at $\sqrt{s_{pp}}$=5.02 TeV\cite{kd17}. More
than three decades ago W and Z bosons were observed at CERN via 
proton-antiproton experiments at $\sqrt{s_{pp}}$=540 GeV\cite{CERN83}. CMS 
experiments on electroweak boson production via relativistic heavy ion 
collisions (RHIC) are related to our present research\cite{CMS11}. 

Our present estimate of the production of Z$^a$ bosons, with $a$ the component 
of the vector Z boson, via Pb-Pb collisions is 
motivated by the fact that Z bosons have little interaction with the nuclear
medium, and by ALICE experiments that measured Z bosons production in  
Pb-Pb collisions\cite{ALICE17} and in p-Pb collisions\cite{ALICE16} at 
$\sqrt{s_{NN}}$=5.02 TeV. The estimate of Z boson production via
Pb-Pb collisions make use of Ref\cite{klm14}, which was based estimates of
heavy quark state production in p-p collisions\cite{klm11}. Note that when
the final calculation and results are presented in Secs 3,4 the momentum
$p^a \rightarrow p_Z$, the momentum of the Z boson produced by Pb-Pb
collisions at $\sqrt{s_{pp}}$=5.02 TeV, and $Z^a \rightarrow Z$, a Z boson.

  Our present work is also related to an estimate of  $\Psi(2S)$ to $J/\Psi$
decay to $\pi$ mesons\cite{kzms17} except the quarks have a vertex with Z 
bosons rather than pions and there is no gluon-Z vertex . Also it was 
shown\cite{lsk09} that the $\Psi(2S)$ state is pproximately 50\%-50\% mixture 
of a standard charmonium and hybrid charmonium state:
\beq
\label{hybrid}
        |\Psi(2S)>&\simeq& -0.7 |c\bar{c}(2S)>+\sqrt{1-0.5}|c\bar{c}g(2S)> \; ,
\eeq
whle the $J/\Psi$ is essentially a standard $q \bar{q}$ state $|J/\Psi(1S)>
\simeq |c\bar{c}(1S)>$, which we use in our estimate of Z$^a$ boson production
via $\Psi(2S) \rightarrow J/\Psi(1S) +Z^a$. Having a hybrid component, 
$c\bar{c}g$, is important for Z boson production from $\Psi(2S)$ decay as the
active gluon component of $\Psi(2S)$ produces a Z boson, as shown in Figure 2 
(Section 2).

\newpage

\section{$J/\Psi$ + Z production in p-p collisions
 with $\sqrt{s_{pp}}$ 5.02 TeV}

The  cross section for $pp\rightarrow \Psi(2S) \rightarrow J/\Psi(1S) +Z^a$
in terms of $f_q$\cite{CTEQ6,klm11}, the quark distribution function, is
\beq
\label{3}
    \sigma _{pp\rightarrow \Psi(2S) \rightarrow J/\Psi(1S)+Z^a} 
&=&   f_q(\bar{x}(y),2m)f_q(a/\bar{x}(y),2m) 
 \sigma_{\Psi(2S) \rightarrow J/\Psi(1S)+Z^a}  \; ,
\eeq  
where 
$y$=rapidity, $a= 4m^2/s=3.6 \times 10^{-7}$ and $\bar{x}(y)= 1.058 
x(y)$. We take $y=0$ in the present work, so $\bar{x}(0)\simeq 6.4 \times 
10^{-4}$

For $\sqrt{s}$=5.02 TeV The quark distribution functions 
$f_q$\cite{CTEQ6,klm11} are
\beq
\label{fg} 
    f_q(\bar{x}(0),2m)&\simeq& 82.37 -63582.36 x(0)\simeq 41.6 \nonumber \\
  f_q(a/\bar{x}(0),2m)&\simeq& 82.37-\frac{a}{\bar{x}(0)}\simeq 82.4 \; .
\eeq

Therefore from Eqs(\ref{2},\ref{3},\ref{fg})
\beq
\label{cross-section}
\sigma _{PbPb\rightarrow \Psi(2S) \rightarrow J/\Psi(1S)+Z^a} &\simeq& 
4.46\times 10^{5} \sigma _{pp\rightarrow \Psi(2S) \rightarrow J/\Psi(1S)+Z^a}
\eeq

We use the notation
\beq
\label{cross-section-f}
\sigma _{pp\rightarrow \Psi(2S) \rightarrow J/\Psi(1S)+Z^a} &\equiv&
\sigma_{HHZ}(p)=g^{\mu \nu}(\Pi_{HZ^a}^{\mu \nu}(p) +\Pi_{HHZ^a}^{\mu \nu}(p))
\; ,
\eeq
with $\Pi_{HZ^a}^{\mu \nu}(p), \Pi_{HHZ^a}^{\mu \nu}(p)$ defined below.

\vspace{1cm}

 The normal $|c\bar{c}(2S)>$ component of $\Psi(2S)$ decaying to 
$|c\bar{c}(1S)>$ with Z$^a$ production via quark-Z coupling 
shown in Figure 1. 

\begin{figure}[ht]
\begin{center}
\epsfig{file=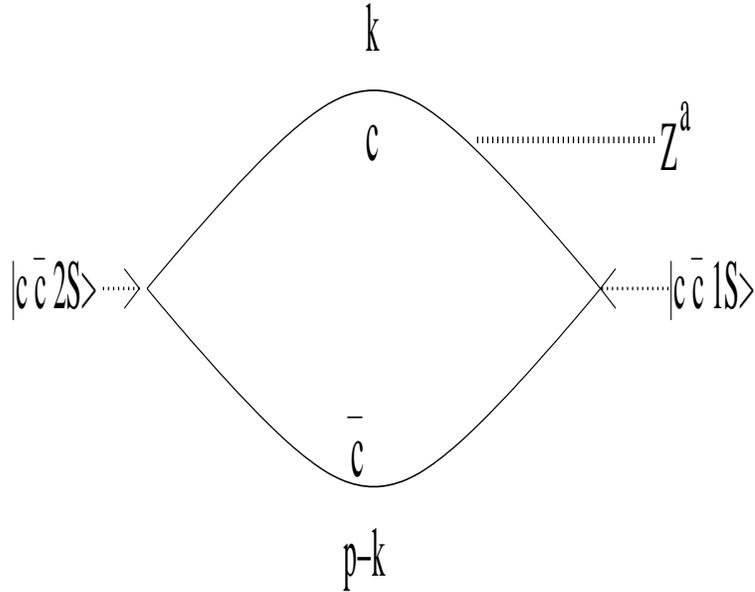,height=8cm,width=10cm}
\end{center}
\caption{$\Psi(2S)$, standard component, to $J/Psi(1S)+Z^a$}
\label{Figure 1}
\end{figure} 
\newpage

In Figure 2 Z$^a$ production with the hybrid  $|c\bar{c}g(2S)>$ component of 
$\Psi(2S)$ it is shown.

\begin{figure}[ht]
\begin{center}
\epsfig{file=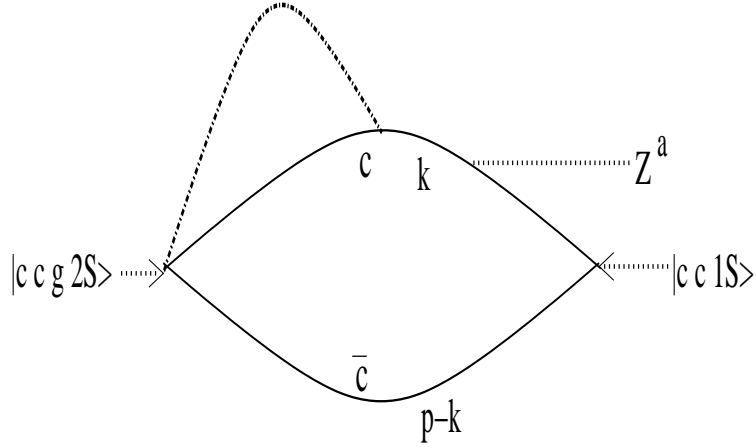,height=6cm,width=10cm}
\end{center}
\caption{ $\Psi(2S)$, hybrid component, to $J/Psi(1S)+Z^a$}
\label{Figure 2}
\end{figure} 

Figure 3 shows the coupling processes needed for Figure 1 and Figure 2.
\vspace{3 cm}
\begin{figure}[ht]
\begin{center}
\epsfig{file=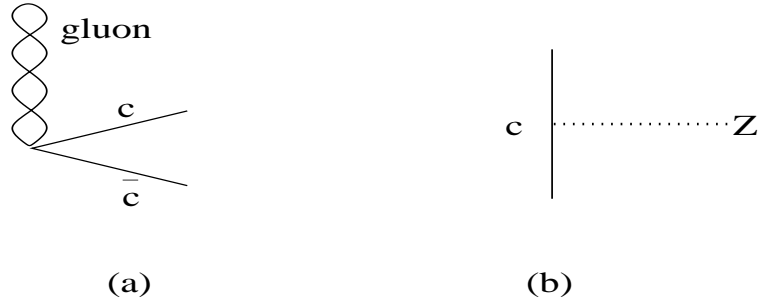,height=4cm,width=10cm}
\end{center}
\caption{(a) gluon-quark coupling, (b) c-Z coupling}
\label{Figure 3}
\end{figure} 

In Figure 3 (a) the operator giving the gluon sigma coupling is
\beq
\label{Gq}
          {\rm gluon-quark\;coupling}&=&\frac{1}{4}S^G_{\kappa \delta}(k)
G^{\kappa \delta}(0) \nonumber \\
          S^G_{\kappa \delta}(k)&=& [\sigma_{\kappa \delta},S(k)]_{+}=
\sigma_{\kappa \delta} S(k)+S(k)\sigma_{\kappa \delta}
 \; ,
\eeq

with $\sigma_{\kappa \delta} =i(\gamma_\kappa \gamma_\delta - g^{\kappa \delta})$ 
and $G^{\kappa \delta}$ is the gluon field.

\vspace{3mm}
In Figure 3 (b), with $Z^a$ the $a$ component of the vector Z boson and 
defining $g_c\equiv g^V_c \simeq 0.25$ \cite{PPB}, the $c c Z^\alpha$ coupling 
is\cite{wh97}
\beq
\label{SZ}
 S_{Z^a}&=&\gamma^a(g_c-g^A_c\gamma^5) 
\eeq
As shown in subsection 3.2 the $\gamma^a \gamma^5$ term does not
contribute to $\sigma_{HHZ}(p)$, so we define  $g^A_c=g_c$.
\newpage

\section{$ \Psi(2S)$ decay to $J/\Psi$ + Z$^a$ }

In this section we estimate the decay of the  $|\Psi(2S)>$ decay to 
$|J/\Psi(1S)> + Z^a$ for both the standard and hybrid components of $|\Psi(2S)>$
as shown in Figures 1 and 2.

\subsection{$\Psi(2S)$  decay to $J/\Psi$ + Z$^a$ via standard 
component of  $\Psi(2S)$}

As in Ref\cite{kzms17} the correlator corresponding to Figure 1 is
\beq
\label{PiHZ}
    \Pi_{HZ^a}^{\mu \nu}(p)&=& \sum_{a b}g^2 \int\frac{d^4k}{(2\pi)^4}
 Tr[S(k)\gamma^\mu S_{Z^a} S(p-k)\gamma^\nu ] \; ,
\eeq
where the quark propagator $S(k)=(\not{k}+M)/(k^2-M^2)$, $M$ is the
mass of a charm quark ($M_c$), $\not{k}=\sum_\mu k^\mu \gamma^\mu$, and 
$g^2=4\pi \alpha_s \simeq 1.49$\cite{PPB}. Since $Tr[S(k)\gamma^\mu S_{Z^a}(k) 
S(p-k)\gamma^\nu]$ is independent of color $\sum_{a b} = 3$.

Thus the correlator for $\Psi(2S)_{normal}$ decay to  $J/\Psi(1S) +Z$ is 
\beq
\label{psi(2S)-Jpsi}
  \Pi_{HZ^a}^{\mu \nu} (p)&=& 3 g^2 g_c \int\frac{d^4k}
{(2\pi)^4} Tr[S(k) \gamma^\mu\gamma^a(1-\gamma^5) S(p-k)\gamma^\nu] \; .
\eeq

The trace in Eq(\ref{psi(2S)-Jpsi}) is
\beq
\label{trace-standard}
   Tr[S(k) \gamma^\mu\gamma^a(1-\gamma^5) S(p-k)\gamma^\nu] &=&
\frac{Tr[(\not{k}+M)\gamma^\mu \gamma^a (1-\gamma^5)[(\not{p}-\not{k})+M)
\gamma^\nu]}{(k^2-M^2)[(k-p)^2-M^2]} \; .
\eeq

  Using the fact that the trace of an odd number of $\gamma$s vanish and
$Tr[\gamma^5 \gamma^\alpha  \gamma^\beta  \gamma^\delta  \gamma^\lambda]=
-4i \epsilon^{\alpha \beta \delta \lambda}$
\beq
\label{trace-standard-2}
 Tr[(\not{k}+M)\gamma^\mu \gamma^a (1-\gamma^5)[(\not{p}-\not{k})+M)\gamma^\nu]
&=& 4M[p_\nu g^{a \mu}+p_a g^{\mu \nu}-p_\mu g^{a\nu}+i\epsilon^{\mu a \alpha
\nu} +2k_\mu g^{a \nu} \nonumber \\
&&-2k_a g^{\mu \nu}-2ik_\alpha \epsilon^{\alpha \mu a \nu}]
\; .
\eeq

Therefore
\beq 
\label{PiHZ-psi(2S)-Jpsi}
  \Pi_{HZ^a}^{\mu \nu}(p)&=&12 g^2 g_c M \int \frac{d^4k}{(4\pi)^4}
\frac{p_\nu g^{a \mu}+p_a g^{\mu \nu}-p_\mu g^{a\nu}+ip_\alpha 
\epsilon^{\mu a \alpha \nu} +2k_\mu g^{a \nu}
-2k_a g^{\mu \nu}-2ik_\alpha \epsilon^{\alpha \mu a \nu}}
{(k^2-M^2)[(k-p)^2-M^2]} \; .
\eeq

Using
\beq
\label{kintegrals}
\int \frac{d^4k}{(4\pi)^4}\frac{1}{(k^2-M^2)[(k-p)^2-M^2]}&=&
\frac{(2M^2-p^2/2)}{(4\pi)^2}I_0(p) \nonumber \\
\int \frac{d^4k}{(4\pi)^4}\frac{k^\mu}{(k^2-M^2)[(k-p)^2-M^2]}&=&
\frac{p^\mu (2M^2-p^2/2)}{(4\pi)^2}I_1(p) \; ,
\eeq
one finds
\beq
\label{PiHZ-psi(2S)-Jpsi-final}
\Pi_{HZ^a}^{\mu \nu}(p)&=&A M\frac{2M^2-p^2/2}{(4\pi)^2}
[(p_\nu g^{a \mu}-p_\mu g^{a \nu}+ip_\alpha \epsilon^{\mu a \alpha \nu})I_0(p)
+(2p_\mu g^{a \nu}-2p_a g^{\mu \nu}-2ip_\alpha \epsilon^{\mu a \alpha \nu})
I_1(p)]\nonumber \\
A&=&12g^2g_c
\; ,
\eeq
with
\beq
\label{I0I1I2}
I_0(p)&=&\int_0^1 d\alpha \frac{1}{p^2(\alpha-\alpha^2)-M^2} {\rm \;\;\;\;}
 I_1(p)=\int_0^1 d\alpha \frac{\alpha}{p^2(\alpha-\alpha^2)-M^2} 
\; .
\eeq

\newpage

\subsection{$\Psi(2S)$  decay to $J/\Psi$ + Z$^a$ via hybrid 
component of  $\Psi(2S)$}

The two-point correlator for the hybrid $\Psi(2S)$-$J/\Psi$, corresponding
to Figure 2  without the gluon-Z or quark-Z coupling is\cite{lsk09}
(see Eq(18))
\beq
\label{Psi(2S)-J-Psi + Z}
    \Pi_{HH}^{\mu \nu}(p)&=& \frac{3g^2}{4}\int \frac{d^4 k}{(2 \pi)^4}
Tr \left [[\sigma_{\kappa \delta} S(k)]_{+} \gamma_\lambda
S(p-k) \gamma_\mu \right ] Tr[G^{\nu \lambda}(0) G^{\kappa \delta}(0)]  
\; .
\eeq

 The correlator $\Pi_{HH q\bar{q} Z^a}^{\mu \nu}$, obtained from Figure 2  is 

\beq
\label{psi(2S)-Jpsi-q-barq-z}
 \Pi_{HHZ^a}^{\mu \nu}(p)&=&\frac{3g^2}{4}\int \frac{d^4 k}{(2\pi)^4}
 Tr \left [[\sigma_{\kappa \delta} S(k)]_{+} \gamma_\lambda S_{Z^a}
S(p-k) \gamma_\mu \right ] Tr[G^{\nu \lambda}(0) G^{\kappa \delta}(0)]  
\; .
\eeq

Note that\cite{kpr09} (with  $<G^2> = .476 {\rm \;GeV}^2$) 

\beq
\label{GG} 
  Tr[G^{\nu \lambda}(0) G^{\kappa \delta}(0)]&=&(2\pi)^4\frac{12}{96}
<G^2>(g^{\nu \kappa} g^{\lambda \delta}-g^{\nu \delta} g^{\kappa \lambda}) \\
 \left [\sigma_{\kappa \delta} S(k) \right ]_{+} &=& i[-2g^{\kappa \delta}
(\not{k}+M)+2M\gamma^\kappa \gamma^\delta +k_\alpha(\gamma^\kappa 
\gamma^\delta \gamma^\alpha +
\gamma^\alpha \gamma^\kappa \gamma^\delta )]/(k^2-M^2) \; .
\eeq

Therefore,
\beq
\label{ZZ}
 \Pi_{HHZ^a}^{\mu \nu}(p)&=&B \int\frac{d^4 k}{(2 \pi)^4}
\frac{i}{(k^2-M^2)[(p-k)^2-M^2]}(g^{\nu \kappa} g^{\lambda \delta}
-g^{\nu \delta} g^{\lambda \kappa}) Tr^{A1}\nonumber \\
M^2B&=& \frac{3g^2 g_c}{4}(2\pi)^4\frac{12}{96} <G^2>\simeq 25.87 
{\rm \;\;GeV^2\;\;\;with} 
\eeq

\beq
\label{TrA1}
Tr^{A1} &\equiv&Tr[( k_\alpha(\gamma^\kappa \gamma^\delta \gamma^\alpha+
 \gamma^\alpha \gamma^\kappa  \gamma^\delta)\gamma^a(1-\gamma_5)\gamma^\lambda
((\not{p}-\not{k})+M)\gamma^\mu \nonumber \\
&&[-2g^{\kappa \delta}(\not{k}+M)+2M\gamma^\kappa \gamma^\delta] \gamma^\lambda
\gamma^a(1-\gamma_5)((\not{p}-\not{k})+M)\gamma^\mu] \; .
\eeq 

Note that $(g^{\nu \kappa} g^{\lambda \delta}-g^{\nu \delta} g^{\lambda \kappa})
g^{\kappa \delta} =0.0$, so the $g^{\kappa \delta}$ term in Eq(\ref{TrA1}) vanishes.
Therefore from Eq(\ref{TrA1})
\beq
\label{TrA1-1}
 (g^{\nu \kappa} g^{\lambda \delta}-g^{\nu \delta} g^{\lambda \kappa})Tr^{A1} 
&=&2Tr\left[[(k_\alpha(\gamma^\nu \gamma^\lambda \gamma^\alpha+
 \gamma^\alpha \gamma^\nu \gamma^\lambda-2 g^{\lambda \nu}\gamma^\alpha)
\gamma^a(1-\gamma_5)\gamma^\lambda+4M(\gamma^\nu- \gamma^\lambda)\gamma^a
(1-\gamma_5)\right ] \nonumber \\
&&((\not{p}-\not{k})+M)\gamma^\mu] \; .
\eeq 

Since Tr[odd number of $\gamma$ s] =0,
\beq
\label{TrA1-2}
 (g^{\nu \kappa} g^{\lambda \delta}-g^{\nu \delta} g^{\lambda \kappa})Tr^{A1} 
&=&2MTr[(k_\alpha(\gamma^\nu \gamma^\lambda \gamma^\alpha+
 \gamma^\alpha \gamma^\nu \gamma^\lambda-2 g^{\lambda \nu}\gamma^\alpha)
\gamma^a(1-\gamma_5)\gamma^\lambda\gamma^\mu \nonumber \\
&&+4(p_\beta-k_\beta)(\gamma^\nu- \gamma^\lambda)\gamma^a(1-\gamma_5)
\gamma^\beta \gamma^\mu] \; .
\eeq

As in Eq(\ref{trace-standard-2}), using $\epsilon^{\alpha \beta \lambda 
\lambda}=0$ one obtains for the $4-\gamma$ terms
\beq
\label{trace-standard-3}
2M Tr[-2k_\alpha g^{\lambda \nu}\gamma^\alpha\gamma^a(1-\gamma_5)\gamma^\lambda
\gamma^\mu+4(p_\beta-k_\beta)(\gamma^\nu- \gamma^\lambda)\gamma^a(1-\gamma_5)
\gamma^\beta \gamma^\mu] &=& \nonumber
\eeq
\beq
\label{trace-standard-4}
16M[-(k_a g^{\mu \nu}+k_\mu g^{a \nu}-k_\nu g^{a \mu}) +ik_\alpha \epsilon^
{\alpha a \nu \mu})  +2[(p_\mu-k_\mu)g^{\nu a}+(p_a-k_a)g^{\nu \mu}-
(p_\nu-k_\nu)g^{\mu a}-\nonumber \\
(p_\mu-k_\mu)g^{\nu a}-(p_a-k_a)g^{\nu \mu}+(p_\nu-k_\nu)
g^{\mu a}
+2i(p_\beta-k_\beta)(\epsilon^{\nu a \beta \mu}-\epsilon^{\nu a \beta \mu})]
\eeq

For the 6-$\gamma$ terms in Eq(\ref{TrA1-1})
\beq
\label{6gamma}
 Tr[\gamma^\alpha \gamma^\beta \gamma^\delta \gamma^\lambda \gamma^\mu 
\gamma^\nu] &=& 4(g^{\alpha \beta}Tr[\gamma^\delta \gamma^\lambda \gamma^\mu 
\gamma^\nu] +g^{\alpha \delta}Tr[\gamma^\beta \gamma^\lambda \gamma^\mu 
\gamma^\nu]+g^{\alpha \lambda}Tr[\gamma^\beta \gamma^\delta \gamma^\mu 
\gamma^\nu]+\nonumber \\
&&g^{\alpha \mu}Tr[\gamma^\beta \gamma^\delta \gamma^\lambda \gamma^\nu]+ 
g^{\alpha \nu}Tr[\gamma^\beta \gamma^\delta \gamma^\lambda \gamma^\mu])
\eeq
\beq
\label{6gamma5}
 Tr[\gamma^\alpha \gamma^\beta \gamma^\delta \gamma^\lambda \gamma^\mu 
\gamma^\nu \gamma_5] &=&-16i(g^{\alpha \beta}\epsilon^{\delta \lambda \mu
\nu}+g^{\alpha \delta}\epsilon^{\beta \lambda \mu \nu}+g^{\alpha \lambda}
\epsilon^{\beta \delta \mu \nu} +...) \; .
\eeq 
\newpage

Fron Eqs(\ref{TrA1-2},\ref{6gamma},\ref{6gamma5}) and using 
$\epsilon^{\beta \beta \mu \nu}=0$, the 6-$\gamma$ terms in Eq(\ref{TrA1-1}) are
\beq
\label{TrA1-6gamma}
 2MTr[(k_\alpha (\gamma^\nu \gamma^\lambda \gamma^\alpha+
 \gamma^\alpha \gamma^\nu \gamma^\lambda) \gamma^a (1-\gamma_5)
\gamma^\beta \gamma^\mu]&=&4M(3k_ag^{\nu \mu}+k_\mu g^{a \nu}+k_\nu g^{a \mu})
\; .
\eeq

Defining $\Pi_{Z}^{\mu \nu}(p)=\Pi_{HZ^a}^{\mu \nu}(p)+\Pi_{HHZ^a}^{\mu \nu}(p)$, with
$p_Z$ the Z boson momentum, from Eqs(\ref{PiHZ-psi(2S)-Jpsi-final},
\ref{TrA1-1},\ref{TrA1-2},\ref{trace-standard-4},\ref{TrA1-6gamma},
\ref{kintegrals})
\beq
\label{PiZa}
 \Pi_{Z}^{\mu \nu}(p)&=&M\frac{2M^2-p^2/2}{(4\pi)^2}[(A(p_\nu g^{a \mu}-p_\mu 
g^{a \nu}+ip_\alpha \epsilon^{\mu a \alpha \nu})+B(3(p_\mu g^{\nu a}+
p_Z g^{\nu \mu}-p_\nu g^{\mu a}-p_\mu g^{\nu a}
\nonumber\\
&&-p_Z g^{\nu \mu}+p_\lambda g^{\mu a} +p_\lambda g^{\mu a})))I_0(p)
+(A(2p_\mu g^{a \nu}-2p_Z g^{\mu \nu}-2ip_\alpha \epsilon^{\mu a \alpha \nu})
\nonumber \\
&&-32 B(p_Zg^{\nu \mu}+p_\mu g^{a \nu}-p_\nu g^{a \mu}-p_Z g^{\mu \nu}-p_\mu 
g^{a \nu}+p_\mu g^{a \nu}) +ip_\alpha \epsilon^{\alpha a \nu \mu})\nonumber \\
&&-2(p_\mu g^{\nu a}+p_Z g^{\nu \mu}-p_\nu g^{\mu a}+p_\mu) g^{\nu a}
-p_Z) g^{\nu \mu}+p_\nu) g^{\mu a})I_1(p) 
\; .
\eeq

From Eqs(\ref{cross-section-f},\ref{ZZ}), taking the $\mu$ sum 
with $ g^{\mu \nu}$
\beq
\label{cross-section-final}
\sigma_{HHZ}(p)&=&4.46\times 10^{5}M\frac{2M^2-p^2/2}{(4\pi)^2}Bp_Z(I_0(p)
+I_1(p)) \nonumber \\
&&=7.4\times 10^4 GeV^2\times p_Z \frac{2M^2-p^2/2}{M}(I_0(p)+I_1(p)) \; ,
\eeq
with the Z boson momentum $p_Z\simeq 1-3$ MeV, so $p_z \ll p$ as $p\simeq 2-3$
 GeV in our calculation. 
 \section{ Calculation of $\sigma _{pp \rightarrow \Psi(2S)}
   \rightarrow \Psi(1S)+Z^a$ via calculation of
   $\sigma_{HHZ}(p)$ for $p \simeq M_c \simeq $ 1.27 GeV} 

 Carrying out the integrals for $I_0(p), I_1(p)$ shown in Eq(\ref{I0I1I2}) one
 obtains from Eq({\ref{cross-section-final}) the values of $\sigma_{HHZ}(p)$,
   with $p_Z$ = 1, 2, 3 MeV = 0.001, 0.002, 0.003 GeV, which from
   Eq(\ref{cross-section-f}) is the cross section $\sigma _{pp\rightarrow
     \Psi(2S) \rightarrow J/\Psi(1S)+Z^a}$  with the proton-proton energy=5.02
   TeV. shown Figure 4 below, 

\begin{figure}[ht]
\begin{center}  
\epsfig{file=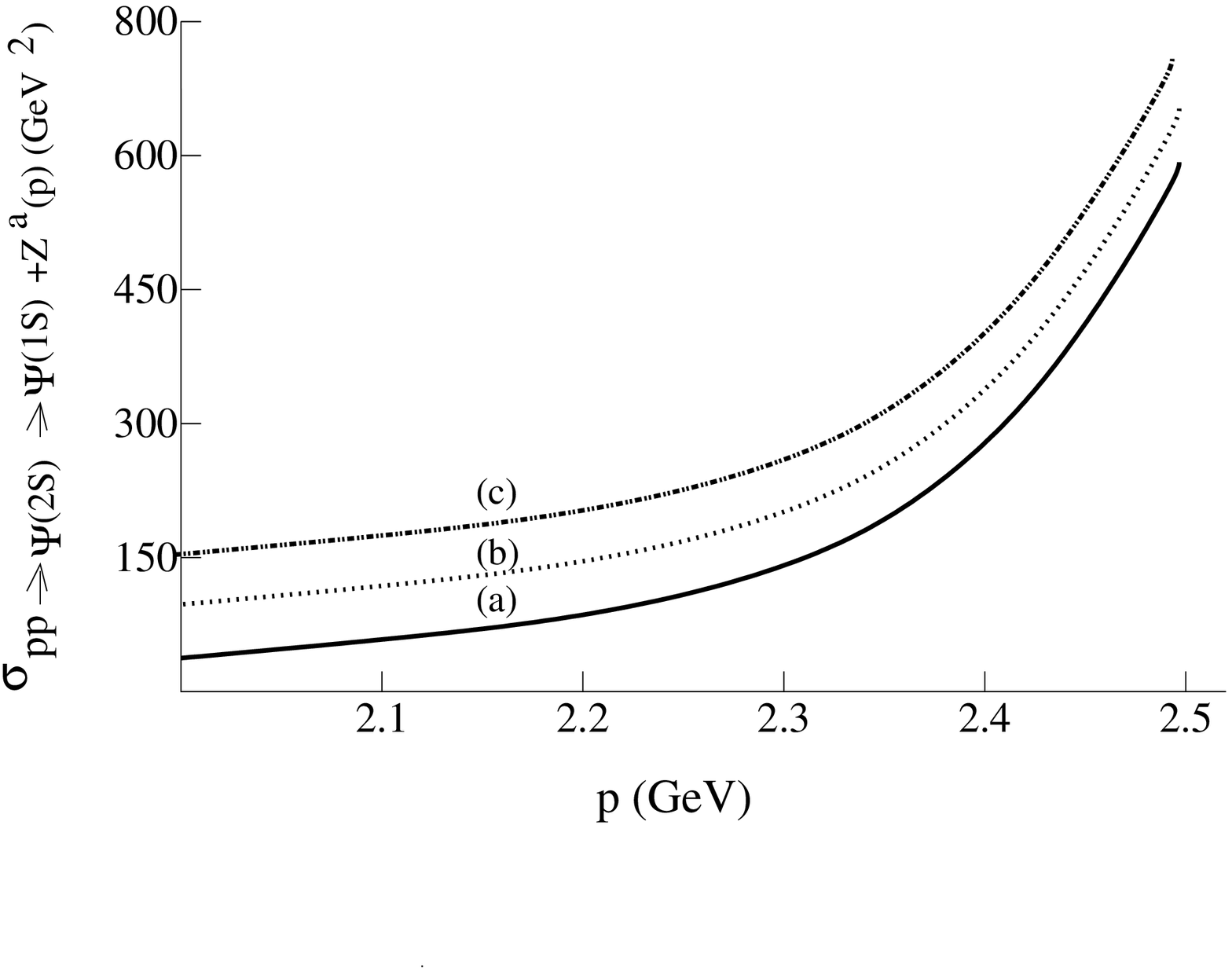,height=6 cm,width=12cm}
\caption{$\sigma _{pp\rightarrow  \Psi(2S) \rightarrow J/\Psi(1S)+Z^a}$ with
  $p_Z$ = (a)1, (b)2, (c) 3 MeV}
\label{figure 4}
\end{center}
\end{figure}

\section{$J/\Psi$ + Z production in Pb-Pb collisions
  with $\sqrt{s_{pp}}$ 5.02 TeV}

 The cross section for the production of a heavy
quark state $\Phi$ with helicity $\lambda=0$ (for unpolarized  
collisions\cite{klm11}) in the color octet model in Pb-Pb collisions is given 
by\cite{klm14}

\beq
\label{2}
   \sigma_{PbPb\rightarrow \Phi} &=& 
 R^E_{PbPb} N^{PbPb}_{bin} \sigma_{pp\rightarrow \Phi} \; ,
\eeq
with $N^{PbPb}_{bin}$ the number of binary collisions and $R^E_{PbPb}$ the
nuclear modification factor in Pb-Pb collisions.
\newpage

 From\cite{kd17} $R^E_{PbPb} N^{PbPb}_{bin}\simeq 130$. Therefore
\beq
\label{final}
\sigma_{Pb-Pb\rightarrow \Psi(2S) \rightarrow J/\Psi(1S)+Z^a} &\simeq&
 130 \times \sigma _{pp\rightarrow \Psi(2S) \rightarrow J/\Psi(1S)+Z^a}
 \; ,
 \eeq

 or $\sigma _{Pb-Pb\rightarrow \Psi(2S) \rightarrow J/\Psi(1S)+Z^a}$ is
 approximately 130 times the results shown in Figure 4.

\section{Conclusions}

Using the relationship between the cross section $\sigma_{PbPb \rightarrow 
\Psi(2S) \rightarrow J/\Psi(1S)+Z}$ and $\sigma_{pp \rightarrow \Psi(2S) 
\rightarrow J/\Psi(1S)+Z}$ shown in Eq(\ref{2}), and  $ \Psi(2S)$ decay to 
$J/\Psi$ + Z for both the the standard and hybrid components of  
$\Psi(2S)$, the cross section $\sigma_{HHZ}(p) \equiv \sigma_{PbPb \rightarrow 
\Psi(2S) \rightarrow J/\Psi(1S)+Z}$ was estimated for $\sqrt{s_{pp}}$=5.02 TeV
and the Z boson momentum  $p_Z$ =1, 2, and 3 MeV, as shown in the figure. 
This should be useful for the experimental measurement of Z boson production 
via Pb-Pb collisions at $\sqrt{s_{pp}}$=5.02 TeV, although for simplicity we 
assumed that the rapidity=y=0, while current experiments\cite{ALICE17} measure 
 Z boson production via Pb-Pb collisions at $\sqrt{s_{pp}}$=5.02 TeV at large
rapidities.
\vspace{1cm}

\Large

{\bf Acknowledgements}
\normalsize
\vspace{5mm}

Author D.Das. acknowledges the facilities of Saha Institute of Nuclear Physics, 
Kolkata, India. Author L.S.Kisslinger acknowledges support in part as a visitor
at the Los Alamos National Laboratory, Group P25. The authors thank Bijit Singha
for helpful suggestions.

\end{document}